\documentclass{IET-Conf-Paper}

\usepackage{booktabs}
\usepackage{amsmath,amssymb,amsfonts}
\usepackage{graphicx}
\usepackage{array,tabularx}   
\usepackage{xcolor}
\usepackage{url}

\newcolumntype{Y}{>{\raggedright\arraybackslash}X}
\newcolumntype{L}[1]{>{\raggedright\arraybackslash}p{#1}}

\begin{document}

\title{GFM Control Automatic Compliance Toolbox}

\author{Tejas Kunbi\ad{1}, Rahul Rane\ad{1}, Dan Wu\ad{2}, Aleksandra Leki\'{c}\ad{1}\corr}

\address{\add{1}{Faculty of Electrical Engineering, Mathematics and Computer Science, Delft University of Technology, Delft, Netherlands}
\add{2}{Shell Nederland B.V., The Hague, Netherlands}
\email{A.Lekic@tudelft.nl}}

\begin{abstract}
The increasing penetration of converter-interfaced generation reduces system inertia, short-circuit strength, and voltage support. European grid codes, including the ENTSO-E Network Code on Requirements for Generators and the ACER Demand Connection Code, now define disturbance-based validation procedures for grid-forming (GFM) generation and converter-based demand. Commercial RMS tools, however, lack an integrated workflow aligned with these recent requirements. This paper presents an automated Python--PowerFactory compliance toolbox and an RMS test bench for a co-located system comprising a 1 MW GFM photovoltaic (PV) plant, a 0.3 MW GFM battery energy storage system (BESS), and a 1 MW converter-interfaced dynamic load at a 20 kV point of common coupling fed from a 110 kV grid. Extended virtual synchronous machine models and a DCC-oriented load model are tested in batch, with results automatically extracted and plotted for assessment. Results show compliant RMS performance for the slower, electromechanical tests, while the paper identifies which tests require complementary electromagnetic transient (EMT) verification.
\end{abstract}


\maketitle

\section{Introduction}
Converter-interfaced generation is displacing synchronous machines, reducing inertia, voltage support, and short-circuit strength; by 2030 converters are expected to exceed 50\% of installed capacity in Europe \cite{IEA2024,EEA2024}. Grid-forming (GFM) control, typically a virtual synchronous machine (VSM), restores these services by operating the converter as a voltage source with synthetic inertia and damping \cite{GFM_Theory,Barac2021}. The ENTSO-E reports on GFM capability of power park modules (PPMs) \cite{ENTSOE2024TechReq,ENTSOE2024SecondGFMReport} extend the static requirements of NC RfG \cite{ENTSOE2016NCRfG} with disturbance-based tests (phase-angle jumps, voltage steps, rate of change of frequency (RoCoF), loss of the last synchronous generator, and short-circuit ratio (SCR) variation), while the ACER Demand Connection Code (NC DCC) \cite{ACER2023NCDC} covers demand.

With several converters at one point of common coupling (PCC), interactions between their power, voltage, and current loops can cause circulating currents and poor power sharing, and require coordinated tuning of droop and virtual impedance, particularly in weak grids \cite{Rosso2021Review,Currentlimiting}. RMS (phasor) models neglect network electromagnetic and inner-loop dynamics and suit electromechanical phenomena above about 100 ms, i.e., inertial and droop response, power sharing, and islanding. Fast converter-driven phenomena, such as current-limiter transients and control interactions below about 10 ms, require EMT models \cite{Hatziargyriou2021Stability}, and quasi-static network models can misjudge the stability of low-inertia systems \cite{Markovic2021SmallSignal}. Hence RoCoF, islanding, and power-sharing tests suit RMS, whereas phase jumps, fast voltage steps, and faults require EMT verification (Table~\ref{tab:Tests}).

VSM models suitable for RMS simulation exist \cite{Barac2021}, and PowerFactory provides a generic grid-forming converter model \cite{PowerFactoryGFM2022}, but detailed GFM studies are mostly EMT-based, which limits the number of repeated compliance runs, and no automated workflow exists for the new GFM tests. The Energinet Model Test Bench \cite{EnerginetSimTools_MTB} and the EASY-RES Python--PowerFactory workflow \cite{FU2024110164,malamaki2021d6} do not provide a test bench aligned with the latest ENTSO-E GFM requirements.

This paper therefore makes a methodological rather than a control-design contribution: (1) the ENTSO-E GFM and NC DCC tests translated into parameterised test definitions annotated with the simulation domain needed for each criterion; (2) RMS models of GFM PV, GFM BESS, and a converter-interfaced load with DC-link dynamics, droop, virtual impedance, and current limitation; (3) a Python--PowerFactory toolbox that runs the test suite in batch and extracts and plots the results, so that the suite can be repeated after any change of tuning or configuration; and (4) a co-located PV--BESS--load case study with coordinated operation for SCR between 2.9 and 26.

\section{Grid Code Compliance Framework} \label{sec:grid_code}
The framework follows the NC RfG and the ENTSO-E GFM reports for generation, and the NC DCC for demand. The ENTSO-E benchmark set-up comprises (i) a controllable AC voltage source with adjustable amplitude, phase, and frequency; (ii) switchable impedances for SCR variation; and (iii) configurable load and generation. Responses are monitored at each unit's point of connection (PoC) and at the 20 kV PCC. Table~\ref{tab:Tests} lists the eight ENTSO-E tests (E1--E8) and three NC DCC tests (D1--D3); the last column indicates whether the success criteria can be verified with RMS models or require EMT confirmation.
\begin{table*}[!t]
\centering
\caption{ENTSO-E GFM (E1--E8) and ACER NC DCC (D1--D3) test specifications and suitable simulation domain}
\label{tab:Tests}
\footnotesize
\setlength{\tabcolsep}{3pt}
\renewcommand{\arraystretch}{1.}
\begin{tabularx}{\textwidth}{@{}L{0.66cm}L{2.6cm}YYL{1.9cm}@{}}
\toprule
\textbf{Test} & \textbf{Name} & \textbf{Disturbance} & \textbf{Success criteria} & \textbf{Domain} \\
\midrule
\multicolumn{5}{@{}l}{\textit{ENTSO-E grid-forming PPM tests}} \\
E1 & Phase angle step & $\pm 5^\circ$ jumps of the grid voltage angle; $P=1.0$ pu & $\Delta i_{p,\text{peak}} \geq 25\%$ of nominal and 50--70\% of the value from $i_P \approx -\frac{u_{\text{inv}}}{x_{\text{eff}}}\sin\delta$; $\geq 50\%$ of change within 10 ms; settling $\leq 60$ ms & EMT (RMS indicative) \\
E2 & Small voltage step & Grid voltage $1.0 \to 0.95$ pu ($0.9$ pu in co-located system); $P=1.0$ pu & $i_q$ response $<10$ ms; 90\% of expected $i_q^+$ within 30 ms; settling 60 ms & EMT (RMS indicative) \\
E3 & Large voltage dip (LVRT) & Grid dip to 0.2 pu, and solid three-phase fault at PCC, 150 ms; $P=1.0$ pu & Stay connected; $i_q$ response $<10$ ms, 90\% within 30 ms; $|i| \leq i_{\max}$; no post-fault overshoot & EMT \\
E4 & Loss of last synchronous generator (LLSG) & Opening of the grid source (islanding); $P=0.5$ pu & Unit becomes voltage/frequency reference; damping $D > 0.05$; no over/under-voltage; supplies 100\% of load & RMS (EMT for first cycles) \\
E5 & RoCoF, over-frequency & Ramp $+4$ Hz/s, 50$\to$51 Hz, 5 s hold; $P=0.5$ pu & 10--90\% $\Delta P$ within 100 ms; synchronism kept; damped & RMS \\
E6 & RoCoF, under-frequency & Ramp $-4$ Hz/s, 50$\to$49 Hz, 5 s hold & As E5 & RMS \\
E7 & SCR variation & Three-phase fault with fault impedance at PCC; SCR 10 and 2.3--2.7 (individual), 26 and 2.9 (co-located) & Stay connected; stable, damped voltage and current; current limits respected & RMS; EMT for control interactions \\
E8 & RoCoF + phase step & $\pm 4$ Hz/s ramp with simultaneous $+5^\circ$ jump; $P=1.0$ pu & Synchronism kept; 10--90\% $\Delta P$ within 100 ms; no sustained oscillation & RMS + EMT \\
\midrule
\multicolumn{5}{@{}l}{\textit{ACER NC DCC demand tests}} \\
D1 & Load frequency response & Frequency steps to 49.0, 47.5, and 51.5 Hz & Connected without limit in 49--51 Hz; time-limited operation outside (scaled to 2 s); no response within $\pm 0.05$ Hz & RMS \\
D2 & RoCoF withstand & $\pm 4$ Hz/s ramps (within E5/E6, co-located) & No disconnection & RMS \\
D3 & Voltage ride-through & Dips to 0.2 and 0.4 pu & Remain connected; reactive current support; $P \geq 90\%$ of pre-fault after recovery & EMT \\
\bottomrule
\end{tabularx}
\end{table*}

\section{Model Development in PowerFactory}
\begin{figure}[!t]
\centering
\includegraphics[width=\columnwidth]{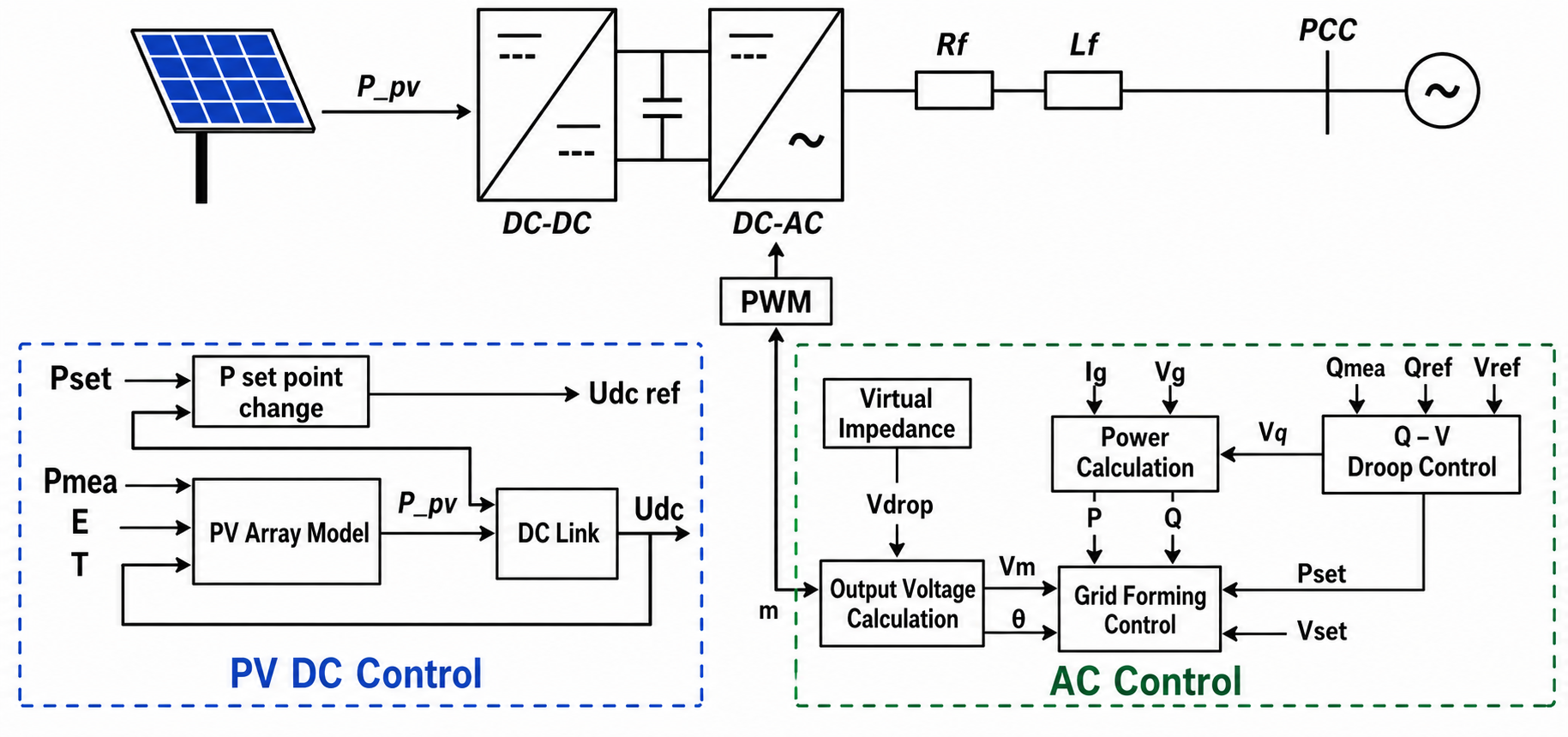}
\caption{GFM PV model structure: PV array and DC-link (MSC side), DC voltage controller, FSM, VSM, AC voltage/Q--V controller, virtual impedance, and current-limited output voltage calculation (GSC side).}
\label{fig:gfm_pv}
\end{figure}

\begin{figure}[!t]
\centering
\includegraphics[width=\columnwidth]{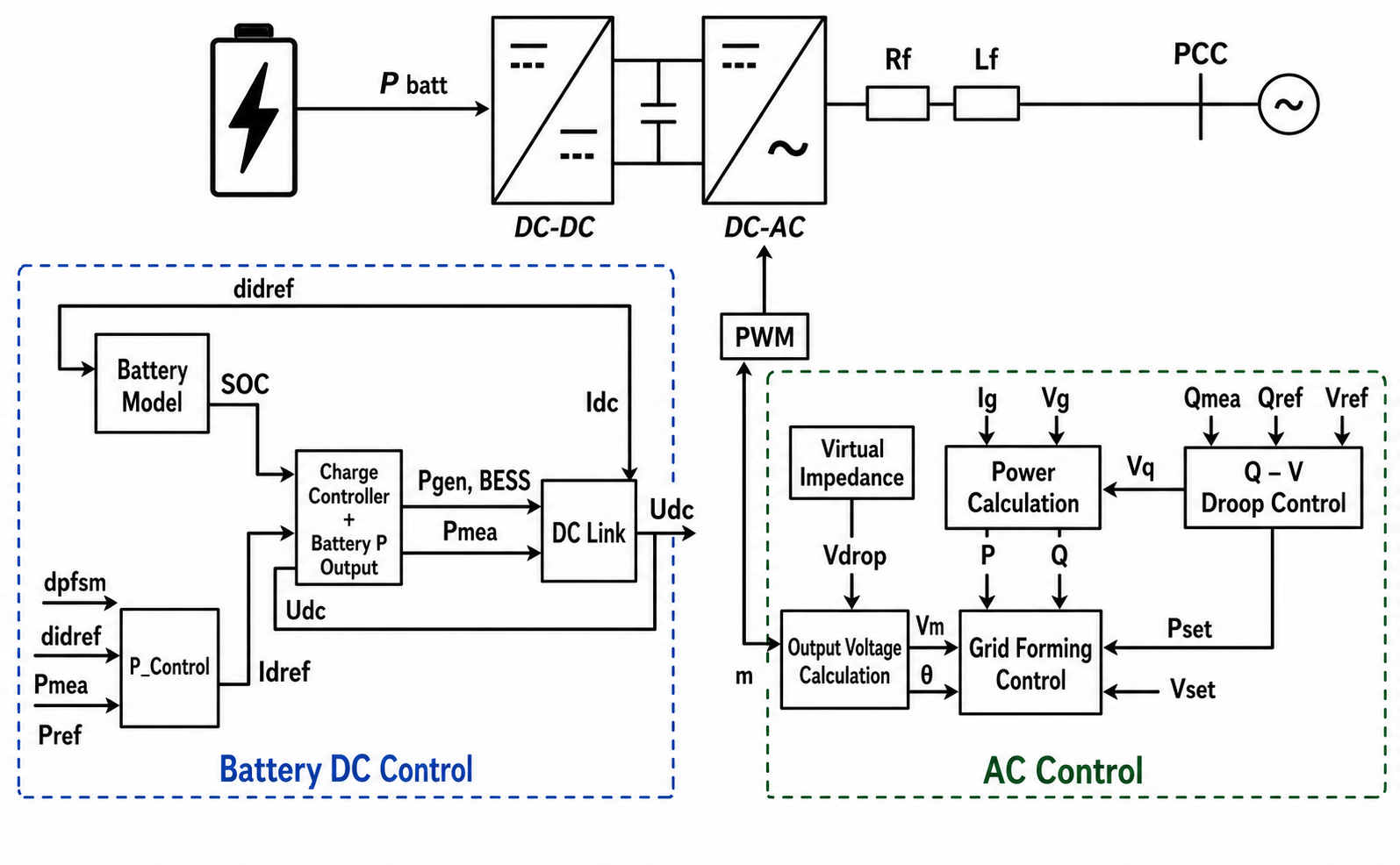}
\caption{GFM BESS model structure: OCV--$R$ battery with SoC-dependent charge controller and DC-link, followed by the same GSC control chain as in Fig.~\ref{fig:gfm_pv}.}
\label{fig:gfm_bess}
\end{figure}

Converters are modeled in DIgSILENT PowerFactory using RMS-averaged representations derived from the standard library models and adapted to the ENTSO-E GFM specifications \cite{ENTSOE2024TechReq}. Each GFM unit is a back-to-back structure in which the machine-side converter (MSC) handles the primary source and the DC link, and the grid-side converter (GSC) forms the PCC voltage using a VSM. Figs.~\ref{fig:gfm_pv} and~\ref{fig:gfm_bess} show the GFM PV and BESS control structures; their blocks are described in Sections~\ref{sec:msc}--\ref{sec:gsc} in the order of the signal flow, from the primary source (left) through the DC link and the outer P--f and Q--V loops to the virtual impedance and current-limited voltage reference (right). Both share the GSC chain and differ in the DC-side source (PV array or OCV- $R$ battery with SoC-dependent charge control) and in their power-control blocks. Fast inner-loop and network electromagnetic dynamics are not represented in the RMS models, which bounds the validity of the results to the domain indicated in Table~\ref{tab:Tests}.

\subsection{Machine-Side Converter (MSC) Control} \label{sec:msc}
\subsubsection{DC-side PV dynamics}
The PV array (left block in Fig.~\ref{fig:gfm_pv}) uses the PowerFactory macro model, in which the array current $I^{\text{PV}}(U_{dc}, E)$ follows from the module MPP, open-circuit, and short-circuit data at standard test conditions, scaled by irradiance $E$, temperature correction factors, and the numbers of series and parallel modules, and $P_{\text{PV}} = U_{dc} I^{\text{PV}}$.

\textit{GFM operation of the PV plant.} A GFM PV inverter can only deliver upward power if the DC side has headroom. In the model, a P-set-point adjustment block shifts the DC-voltage reference with a PI law, $U_{dc,\text{ref}} = (K_{p,u} + K_{i,u}/s)(P_{\text{ref}} - P_{\text{gen,PV}})$, which moves the array along its I--V curve away from the maximum power point and thus curtails it to the dispatched power. The upward headroom is therefore set by the dispatch: the frequency and islanding tests start at 0.5 pu (individual units) or at 0.8 MW of the PV's 1 MW rating (co-located system), under constant irradiance. At 1.0 pu loading the upward response is clamped at rated power (Section~\ref{sec:simulations}). A supervisory MPPT/headroom layer is not implemented. When the headroom is exhausted (or at night), the PV plant's GFM capability is limited to downward power and reactive support, which motivates the BESS-referenced configuration discussed in Section~\ref{sec:coord}.

\subsubsection{DC-side battery dynamics}
The battery (Fig.~\ref{fig:gfm_bess}) is an aggregated OCV--$R$ model scaled to pack level, with $C_r$ cells in series, $C_p$ strings in parallel, and cell capacity $C_c$ (Ah):
\begin{align}
I_{\text{cell}} &= I/C_p, \qquad \frac{d\,SoC}{dt} = -\frac{I_{\text{cell}}}{3600\,C_c}, \\
U_{\text{cell}} &= U_{\max} SoC + U_{\min}(1 - SoC) - I_{\text{cell}} R_{\text{cell}}, \\
V_{\text{term}} &= C_r U_{\text{cell}}.
\end{align}
The pack has $C_r = 31$ series and $C_p = 19$ parallel cells of 40 Ah and is connected at 0.4 kV. The DC-voltage controller keeps $U_{dc}$ close to $V_{\text{term}}$, so that the converter does not force the battery to unrealistic operating points.

\textit{Impact of SoC on GFM behavior.} The SoC acts on the GFM response in two ways. First, it sets the terminal voltage and hence the DC voltage and current margin. Second, a charge controller overrides the active-current reference $i_{d,\text{ref}}$ from the power loop: it forces charging below $SoC_{\min}$ and discharging above $SoC_{\max}$, and passes $i_{d,\text{ref}}$ unchanged in between. Near these limits, the VSM's upward or downward range is clipped, reducing its effective inertia and droop contribution in that direction. Because the tests last less than a minute, the SoC changes only marginally during a test. The current logic, however, does not always switch from charging to discharging when frequency support is needed; a priority rule between SoC management and frequency support is left for future work.

\subsubsection{DC link}
The DC link buffers the mismatch between the source power $P_{in}$ and the converter output $P_{out}$:
\begin{equation}
C U_{dc} \frac{dU_{dc}}{dt} = P_{in} - P_{out}.
\end{equation}
It is parameterised by the capacitor time constant $T_{\text{cap}} = C U_{dc,n}^2/P_{\text{rated}}$, so that the stored energy is $E = \tfrac{1}{2} P_{\text{rated}} T_{\text{cap}}$. Table~\ref{tab:Tcap} lists the values used. These values do not represent a physical inverter capacitor, whose $T_{\text{cap}}$ is in the order of tens of milliseconds at most; they represent an \emph{equivalent} DC-side energy buffer that aggregates the DC capacitor and the fast-acting headroom of the DC source (PV de-loading, or DC-coupled storage). This is a modeling assumption that isolates the GSC GFM behavior from DC-side energy limitations. The value of 2 s was chosen to limit DC voltage deviations to about 5--10\% during transients; smaller values tighten the coupling between the DC-voltage loop and the P--f loop and can induce poorly damped low-frequency oscillations. With a physical capacitor only, the inertial energy must instead come from the primary source within the DC-voltage controller bandwidth; a sensitivity study with physical values is left for future work.

\begin{table}[!t]
\centering
\caption{Equivalent DC-side energy buffer}
\label{tab:Tcap}
\begin{tabular}{@{}lccc@{}}
\toprule
Plant & $P_{\text{rated}}$ (MW) & $T_{\text{cap}}$ (s) & $E$ (kJ) \\
\midrule
PV   & 1.0 & 2.0 & 1000 \\
BESS & 0.3 & 1.5 & 225 \\
\bottomrule
\end{tabular}
\end{table}

\subsubsection{DC voltage controller} \label{sec:dcctrl}
The DC voltage controller balances the DC-link energy by adjusting the inverter active-power set point:
\begin{equation}
\begin{aligned}
p_{\text{set,inv}} = p_{\text{gen}} &+ K_{p,dc} \Delta U_{dc} + K_{i,dc} \!\int\! \Delta U_{dc}\, dt \\
&+ K_{d,dc} \frac{d\Delta U_{dc}}{dt},
\end{aligned}
\end{equation}
where $p_{\text{gen}}$ is a feed-forward of the DC source power and $\Delta U_{dc} = U_{dc,\text{ref}} - U_{dc}$. The gains were initialized using the library defaults and adjusted to achieve a well-damped DC voltage with limited overshoot.

\subsection{Grid-Side Converter (GSC) Control} \label{sec:gsc}
The GSC uses a VSM with the cascaded blocks shown on the right-hand side of Figs.~\ref{fig:gfm_pv} and~\ref{fig:gfm_bess} \cite{Barac2021}. All quantities are per unit on the converter rating unless stated otherwise.

\subsubsection{Frequency-sensitive mode (FSM)}
The FSM block adds the NC RfG frequency-sensitive response $\Delta P_{\text{FSM}}$: a continuous droop term and the limited frequency-sensitive modes for over- and under-frequency (LFSM-O/U, droops $s_3$, $s_4$), each limited, rate-limited, and filtered.

\subsubsection{VSM}
The VSM block computes the internal frequency $\omega$ and angle $\theta$ from the swing equation
\begin{equation}
T_a \frac{d\omega}{dt} = p_{\text{set}} - p_{\text{mea}} - D_p (\omega - \omega_{\text{set}}), \quad \frac{d\theta}{dt} = \omega_b \omega,
\end{equation}
where $T_a = 2H$ is the acceleration time constant, $p_{\text{set}}$ includes $\Delta P_{\text{FSM}}$, $p_{\text{mea}}$ is the measured output power, and $D_p$ is the damping coefficient. $T_a$ is selected from the required power response to a given RoCoF, $T_a = \Delta p\, f_{\text{rated}}/\text{RoCoF}$. A 10\% P--f droop, within the 2--12\% range of ENTSO-E, gives $\Delta p = 0.2$ pu for a 1 Hz deviation; with RoCoF $= 4$ Hz/s, this yields $T_a = 0.2 \cdot 50/4 = 2.5$ s, used for both units.

\subsubsection{AC voltage and Q--V control}
The AC voltage controller adjusts the internal voltage magnitude:
\begin{equation}
u_{\text{set}}(s) = \frac{K_v}{1 + s T_v} \left[(u_{\text{ref}} - u) + \Delta u_{\text{OEL}} + \Delta u_{\text{UEL}}\right],
\end{equation}
where $K_v$ and $T_v$ are the gain and filter time constant, and $\Delta u_{\text{OEL}}$, $\Delta u_{\text{UEL}}$ are over- and under-excitation limiter corrections. A Q--V droop loop is added for reactive power sharing and voltage support:
\begin{equation}
V_m(s) = \left[D_q(V_{\text{ref}} - V) + (Q^* - Q)\right] \left(K_{p,q} + \tfrac{K_{i,q}}{s}\right),
\end{equation}
where $V_m$ is the internal EMF magnitude reference, $D_q$ the Q--V droop, $Q^*$ the reactive power set point, and $K_{p,q}$, $K_{i,q}$ the PI gains.

\subsubsection{Virtual impedance (VI)}
The VI block emulates a series impedance between the internal EMF and the terminal:
\begin{equation}
\Delta v_{d} = r_{vi} i_{d} - x_{vi} i_{q}, \qquad \Delta v_{q} = r_{vi} i_{q} + x_{vi} i_{d},
\end{equation}
where $r_{vi}$, $x_{vi}$ are the virtual resistance and reactance and $i_d$, $i_q$ the converter currents in the VSM reference frame. During overcurrent the impedance increases with the current excess:
\begin{equation}
z_{vi} =
\begin{cases}
r_{vi} + j x_{vi}, & |i| \leq i_{\text{lim}} \\
r_{vi} + j x_{vi} + (k_{pr} + j k_{px})\Delta i, & |i| > i_{\text{lim}}
\end{cases}
\end{equation}
with $\Delta i = |i| - i_{\text{lim}}$, which limits fault current while preserving the voltage-source behavior and adds damping \cite{Currentlimiting}.

\subsubsection{Output voltage and current limitation}
The converter voltage is the internal VSM voltage minus the VI drop. The drop between this reference and the filtered terminal voltage is limited in magnitude, preserving its angle:
\begin{equation}
u_{\text{drop,max}} = |Z_{\text{series}}|\, i_{\max}, \qquad u_{\text{conv}} = u_{\text{term}} + u_{\text{drop,lim}},
\end{equation}
where $i_{\max}$ is the maximum converter current (1.5 pu here) and $Z_{\text{series}}$ is the virtual series impedance. Finally, $|u_{\text{conv}}| \leq u_{\text{conv,max}}$ is imposed, and first-order filters on magnitude and angle ensure smooth transitions near the limits.

\subsection{Dynamic Load Model}
Conventional ZIP and WECC composite load models \cite{WECC2021CLM} lack autonomous control of frequency or reactive power. A converter-interfaced GFL load representing aggregated industrial demand \cite{DatacentreDERinvert} was therefore implemented as a user-defined composite model with a PLL, an inner current controller, and outer power loops. With $P_{\text{ref}}$ the consumed active power and $Q_{\text{ref}}$ the injected reactive power,
\begin{equation}
\begin{aligned}
P_{\text{ref}} &= P_{\text{nom}} - K_{Pf} |\Delta f|, \\
Q_{\text{ref}} &= Q_{\text{nom}} - K_{QV} (V - V_{\text{ref}}),
\end{aligned}
\end{equation}
so that consumption is reduced for both under- and over-frequency (5\% droop), and reactive injection increases at low voltage. Voltage- and frequency-dependent tripping is represented by participation multipliers, with inner thresholds $V_{L1}$, $V_{H1}$ and outer thresholds $V_{L0}$, $V_{H0}$:
\begin{equation}
M_V =
\begin{cases}
0, & V < V_{L0} \text{ or } V > V_{H0}, \\
\frac{V - V_{L0}}{V_{L1} - V_{L0}}, & V_{L0} \le V \le V_{L1}, \\
1, & V_{L1} < V < V_{H1}, \\
\frac{V_{H0} - V}{V_{H0} - V_{H1}}, & V_{H1} \le V \le V_{H0},
\end{cases}
\end{equation}
\begin{equation}
M_f =
\begin{cases}
0, & f \le f_{\min} \text{ or } f \ge f_{\max}, \\
1 - K_f |f - f_{\text{nom}}|, & f_{\min} < f < f_{\max},
\end{cases}
\end{equation}
and $M_{\text{total}} = M_V M_f$ scales the current reference. The load is rated 1 MW at 0.69 kV with $\pm 48\%$ reactive capability; an LVRT function keeps it connected during short dips.

\subsection{Network Model}
\begin{figure}[!t]
\centering
\includegraphics[width=\linewidth]{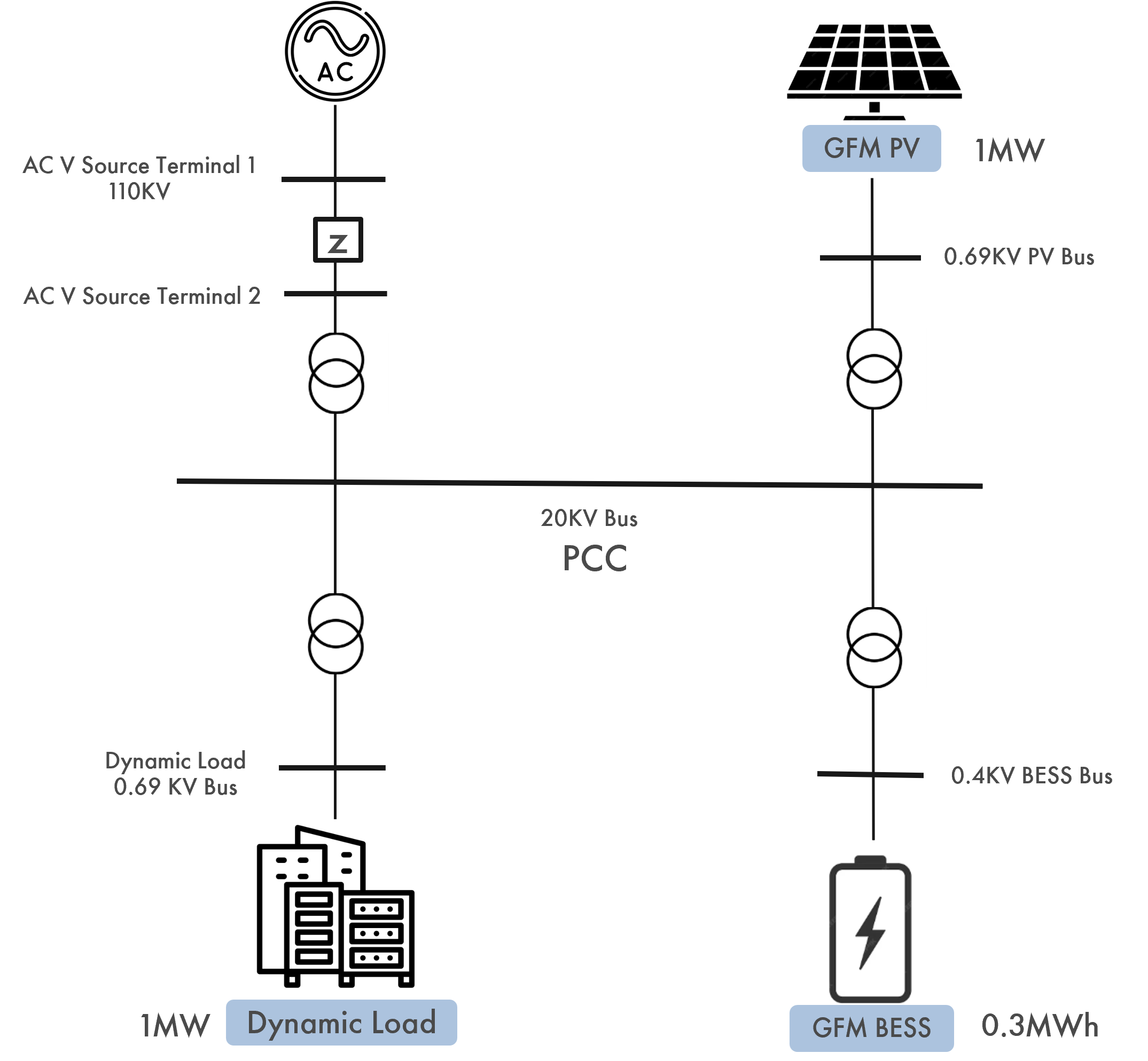}
\caption{Co-located test system: 1 MW GFM PV, 0.3 MW GFM BESS, and 1 MW dynamic load at the 20 kV PCC, connected to a 110 kV source through a transformer and a variable R--L grid impedance.}
\label{fig:colocated_systm}
\end{figure}

The co-located system in Fig.~\ref{fig:colocated_systm} connects the PV plant, the BESS, and the dynamic load to a 20 kV PCC, which is fed from a 110 kV AC voltage source through a two-winding transformer and an R--L branch representing the Th\'evenin grid impedance. Varying this branch ($X/R =10$) sets the SCR at the PCC between 2.3 and 26. Frequency and RoCoF events are applied through the frequency controller of the AC voltage source; voltage dips and phase jumps are applied through events on its magnitude and angle; and faults are applied directly at the PCC.

\subsection{Tuning Procedure and Multi-Converter Coordination} \label{sec:coord}
Tuning was performed in two stages.

\textit{Stage 1 -- individual units.} Each converter is tested alone against the AC source, with all other elements out of service, so that its response reflects only the grid disturbance. A single parameter set per converter is used for all tests and SCR values, so that differences in results arise solely from the disturbance. With the initial gains, the units became oscillatory at the lowest SCR due to coupling between the Q--V droop, the AC voltage loop, and the virtual impedance. In deep sags, a high Q--V proportional gain drove the converter into its current limit; the voltage loop then over-corrected, and post-fault overshoots of 1.2--1.4 pu or loss of ride-through occurred. Stability was obtained by (i) taking the lowest SCR as the design case; (ii) sizing the effective series impedance so that the worst-case fault current stays within the limit, $|Z_{\text{eff}}| \geq \Delta V_{\max}/i_{\max}$, which gives $x_{vi} \approx 0.2$--$0.4$ pu, with a small $r_{vi}$ for damping \cite{Currentlimiting}; (iii) reducing the voltage-loop bandwidth and tuning the Q--V loop within the capability limit $P^2 + Q^2 \leq (V_t\, i_{\max})^2$; and (iv) adding anti-windup to all integrators that interact with the current limiter. The resulting single set per converter (Table~\ref{tab:gfm_parameters}) was stable for SCR from about 2.3 to 10, for shallow and deep dips and for phase jumps.

\textit{Stage 2 -- co-located system.} Connecting the units in parallel lowers the equivalent impedance seen by each converter, increases the short-circuit power at the PCC, and introduces power-sharing and circulating-current interactions, including with the GFL load. Because the effective impedance at the PCC changes, the virtual impedances of the individual stages made the units too stiff and caused excessive reactive current; $x_{vi}$ was therefore reduced from 0.4 to 0.1 pu and $K_v$ from 3 to 2. The parameters were re-tuned following three rules: (1) \textit{reference unit}: one GFM unit has a stiffer voltage reference and faster Q--V loop, while the other uses a lower droop gain and follows it; (2) \textit{droop design}: P--f and Q--V droops are proportional to unit ratings and sized for the weakest grid (SCR 2.9) to bound voltage and frequency deviations; (3) \textit{VI and current limits}: the virtual impedance provides damping without reducing voltage-control bandwidth, and current limits are coordinated so that both units can reach rated current simultaneously \cite{Rosso2021Review,Currentlimiting}. Table~\ref{tab:gfm_parameters} gives both parameter sets. After retuning, the co-located system was stable for SCR between 2.9 and 10; with a strong grid (SCR 25--30) all co-located tests ran without instability, and the weak-grid case (SCR $=2.9$) was verified with the fault test of Section~\ref{sec:simulations}.

Retuning between the two stages does not contradict the workflow: compliance is a property of a unit within a given network configuration, and the toolbox allows the test suite to be rerun after each change. Parameter selection itself remains manual.

In this study, the PV plant is the reference unit, representing daytime operation with irradiance headroom due to its larger rating. Because the PV primary source is uncontrollable, a BESS-referenced configuration is the robust choice for night-time, cloud transients, or exhausted headroom. Swapping the reference unit only changes the parameter sets in the test definitions; this configuration is left for future work.

\begin{table}[!t]
\centering
\caption{GFM parameters for individual / co-located tests (pu). \\ Parameters not listed are based on the PowerFactory library model.}
\label{tab:gfm_parameters}
\footnotesize
\begin{tabular}{@{}lcc@{}}
\toprule
Parameter & PV & BESS \\
\midrule
Rating (MW) / voltage (kV) & 1.0 / 0.69 & 0.3 / 0.4 \\
Acceleration time constant $T_a$ (s) & 2.5 & 2.5 \\
P--f droop (\%) & 10 & 10 \\
AC voltage gain $K_v$ & 3 / 2 & 3 / 2 \\
Q--V PI gains $K_{p,q}$ / $K_{i,q}$ & 2.5 / 4 & 2 / 4 \\
Virtual reactance $x_{vi}$ & 0.4 / 0.1 & 0.4 / 0.1 \\
Virtual resistance $r_{vi}$ & 0.01 & 0.02 \\
Current limit $i_{\max}$ & 1.5 & 1.5 \\
\bottomrule
\end{tabular}
\end{table}

\section{Toolbox Scripting and Automation}
The automation uses the PowerFactory Python API \cite{python}. An Excel sheet lists, row by row, the study folder, study case, test case, event type, and attributes (time, variable, value), a run flag, and a rank that groups rows for a given test case. For each rank, the script activates the study case, sets all events out of service and activates only those of the current test, updates their attributes, runs the RMS simulation, extracts the predefined signals (P, Q, voltage, frequency, positive-sequence currents) from the result object, and plots them for assessment against the criteria in Table~\ref{tab:Tests}. The same script was used for the PV, BESS, dynamic load, and co-located system, reducing manual setup and post-processing effort by more than 90\%. The current implementation relies on fixed object and variable names in the project, so new assets require adaptation of the mappings.

\section{Simulation Results} \label{sec:simulations}
Three representative ENTSO-E tests are discussed: RoCoF (E5/E6), LLSG (E4), and SCR variation (E7). These are the tests whose criteria can be assessed reliably in the RMS domain (Table~\ref{tab:Tests}); the dynamic load response is included in each, and its NC DCC tests are summarised in Section~\ref{sec:dcc}.

\subsection{RoCoF Response (E5, E6)}
\begin{figure}[!t]
\centering
\includegraphics[width=\columnwidth]{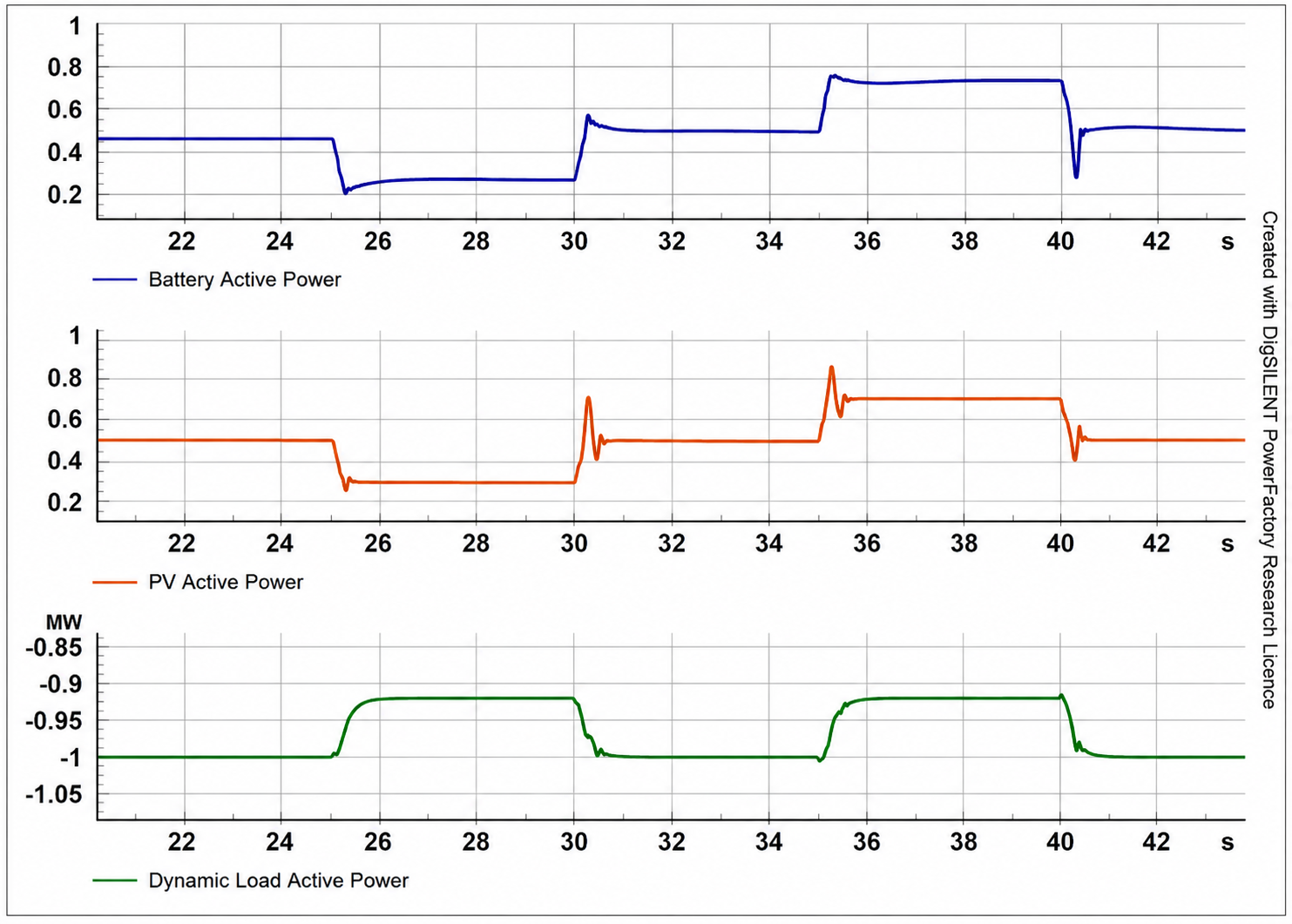}
\caption{Active power response during RoCoF up and down tests (E5, E6).}
\label{fig:rocof_p}
\end{figure}

The PV--BESS system is subjected to $\pm4$ Hz/s frequency ramps with initial conditions $P=0.5$ pu, $Q=0$, and $V_{\text{PoC}}=1.0$ pu. Fig.~\ref{fig:rocof_p} shows the active power response. Synchronism is maintained, and the inertial and droop-based power adjustment is smooth and well damped in both directions. The 10--90\% power change is reached within 100 ms in both directions, and the internal VSM frequency follows the grid with an error below 0.02 Hz. When the test is started at 1.0 pu, the upward response during the under-frequency ramp is clamped at rated power, and the inertial part is not visible; this illustrates the headroom requirement discussed in Section~\ref{sec:msc}. The dynamic load follows its 5\% droop symmetrically without adding oscillations at the PCC.

\subsection{Loss of Last Synchronous Generator (E4)}
\begin{figure}[!t]
\centering
\includegraphics[width=\columnwidth]{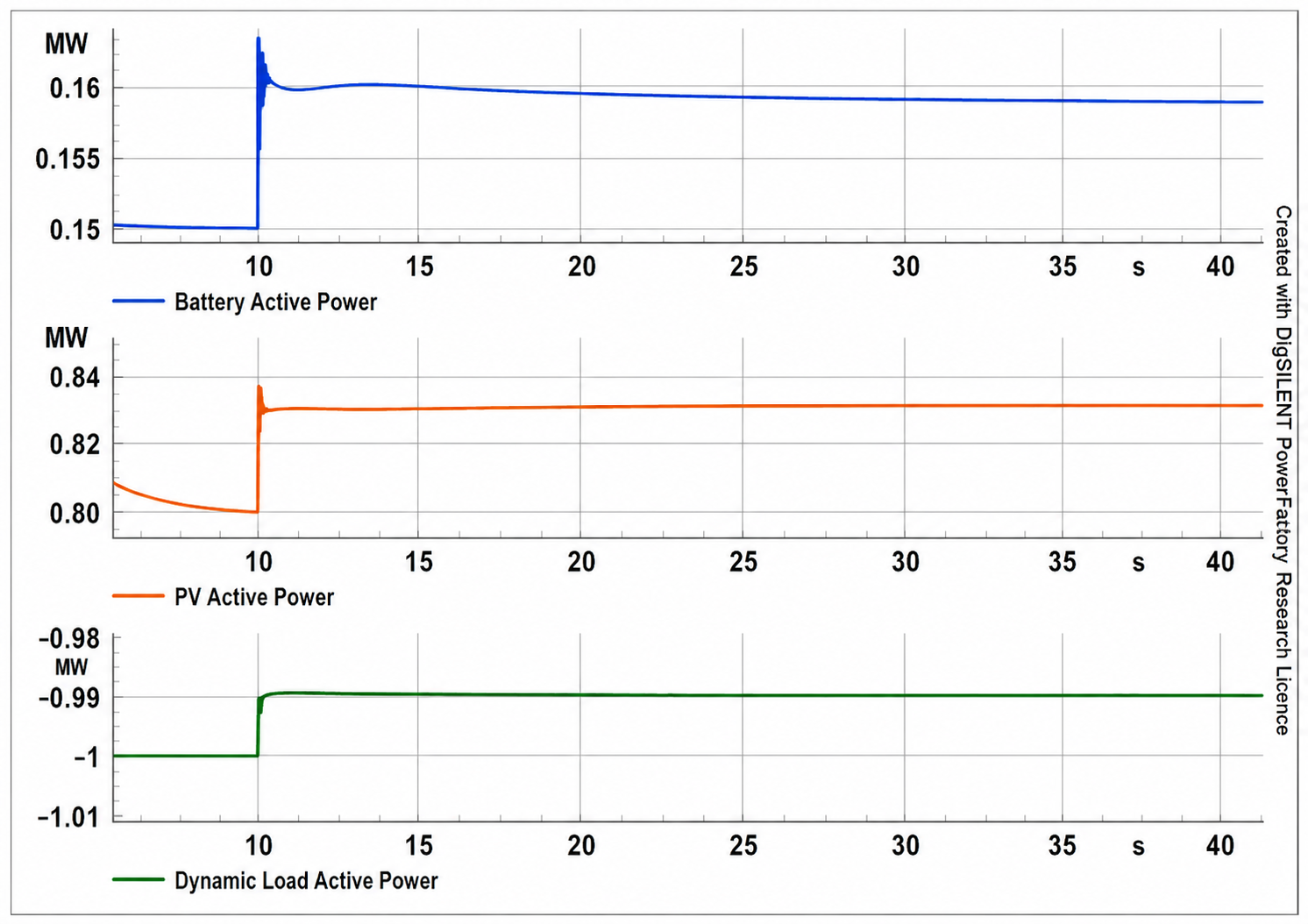}
\caption{Active power of PV, BESS, and load during the LLSG test (E4).}
\label{fig:llsg_power}
\end{figure}

\begin{figure}[!t]
\centering
\includegraphics[width=\columnwidth]{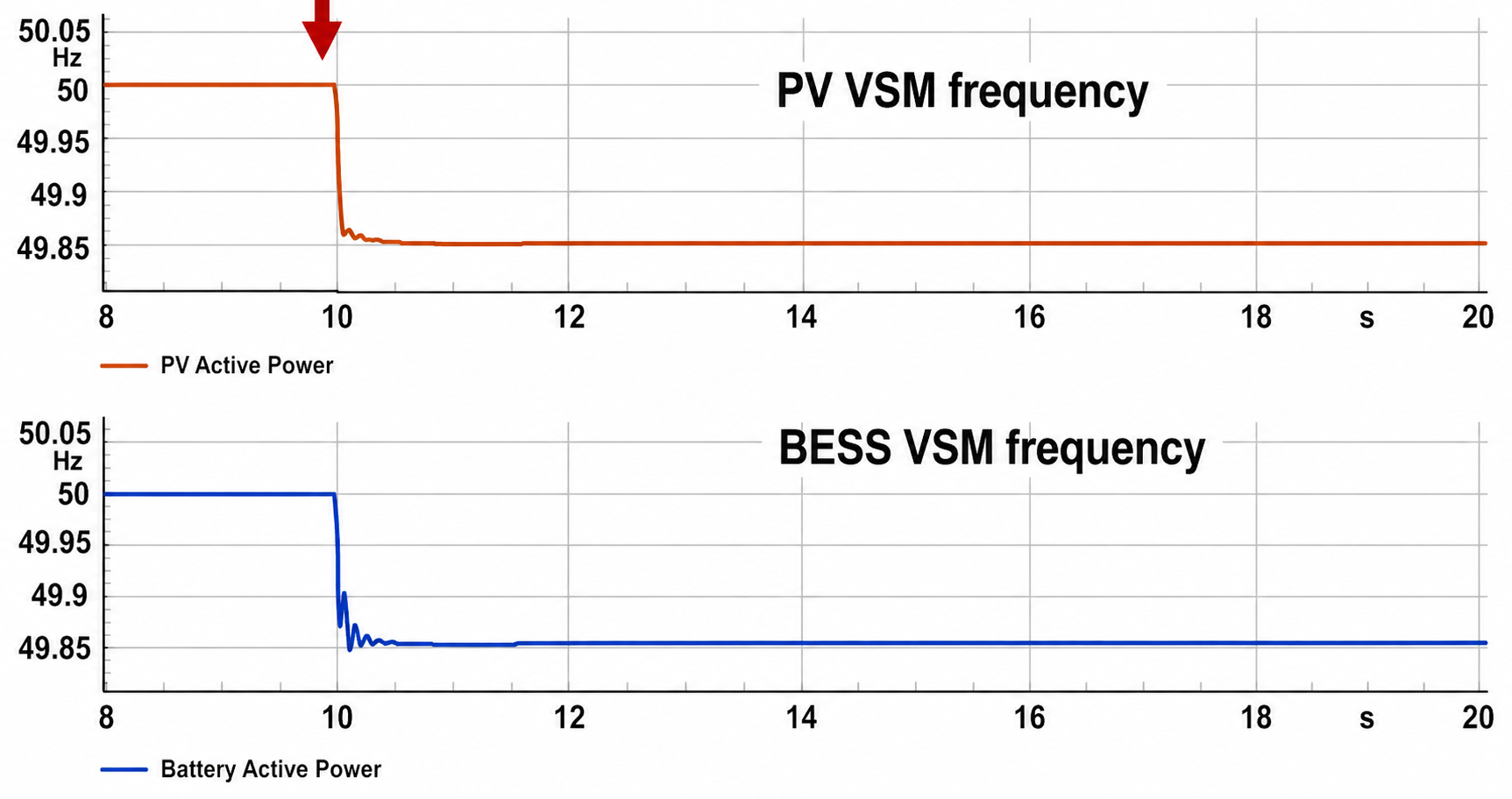}
\caption{PV and BESS frequency during the LLSG test (E4).}
\label{fig:llsg_freq}
\end{figure}

\begin{figure}[!t]
\centering
\includegraphics[width=\columnwidth]{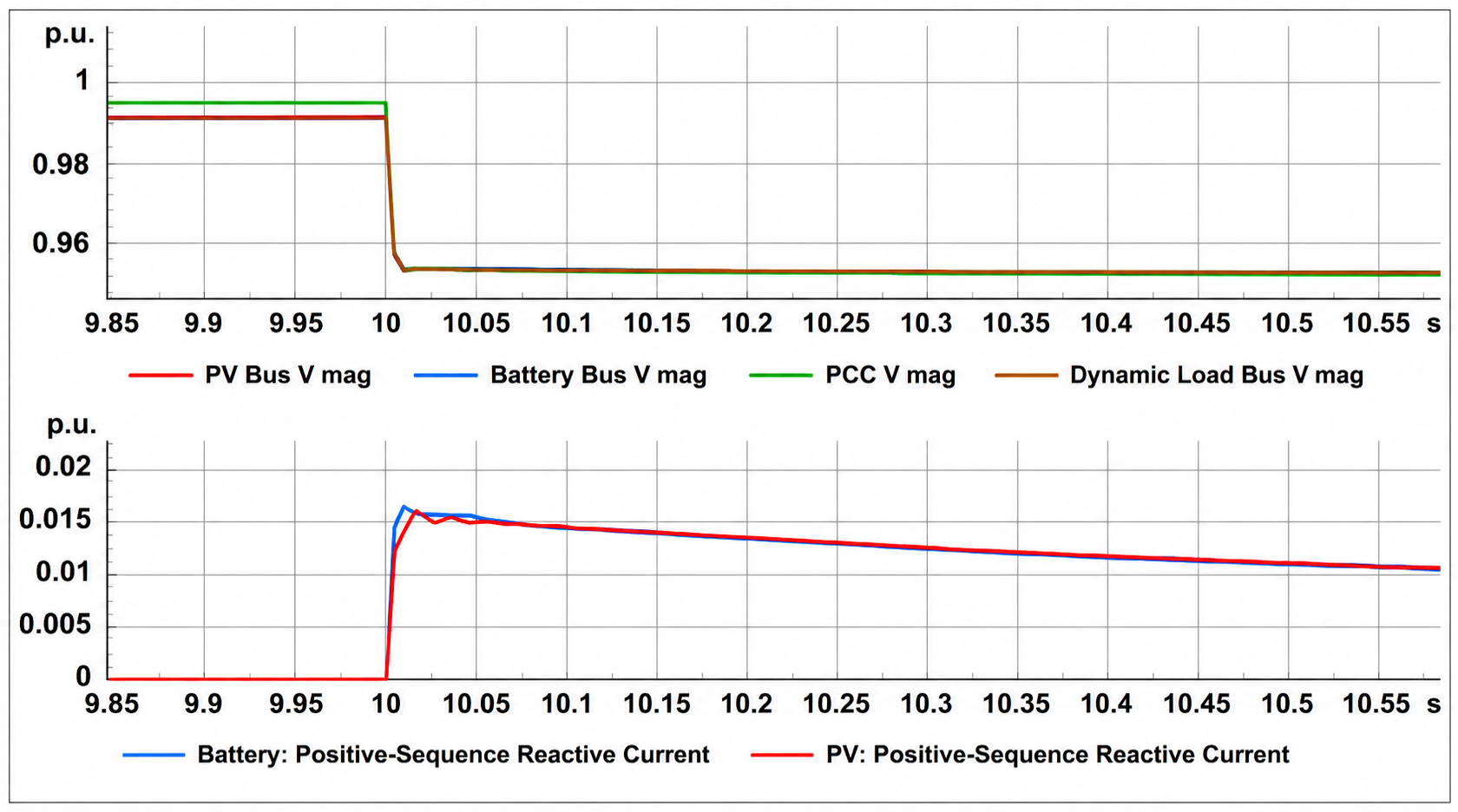}
\caption{PCC voltage and reactive current during the LLSG test (E4).}
\label{fig:llsg_viq}
\end{figure}

The grid is disconnected at $t \approx 10$ s. With a 10\% P--f droop and $T_a = 2.5$ s, the PV and BESS increase their output from about 0.80/0.15 MW to 0.83/0.16 MW and supply the 0.99 MW load (Fig.~\ref{fig:llsg_power}). The frequency of both units (Fig.~\ref{fig:llsg_freq}) dips to about 49.5--49.6 Hz and settles near 49.85 Hz without oscillation. The PCC voltage (Fig.~\ref{fig:llsg_viq}) drops from about 0.99 to 0.955 pu, within 0.9--1.1 pu, with a small reactive power step from both GFM units. The dynamic load follows the network conditions set by the PV reference unit, and stable islanded operation is achieved.


\subsection{SCR Variation (E7)}
\begin{figure}[ht!]
\centering
\includegraphics[width=\columnwidth]{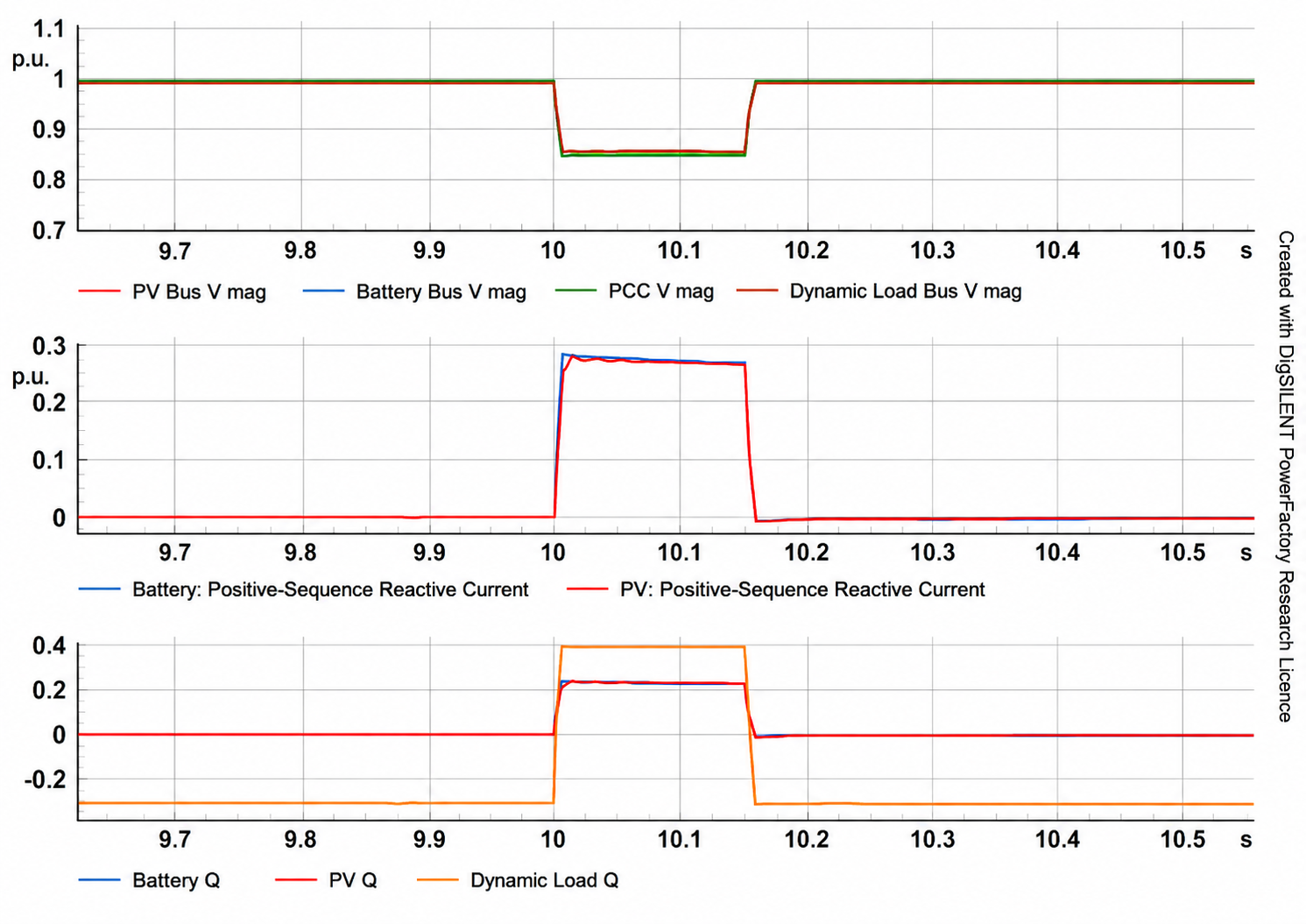}
\caption{Voltage, reactive current, and reactive power in the SCR test, very strong grid (SCR $=26$).}
\label{fig:scr26_qv}
\end{figure}

\begin{figure}[!t]
\centering
\includegraphics[width=\columnwidth]{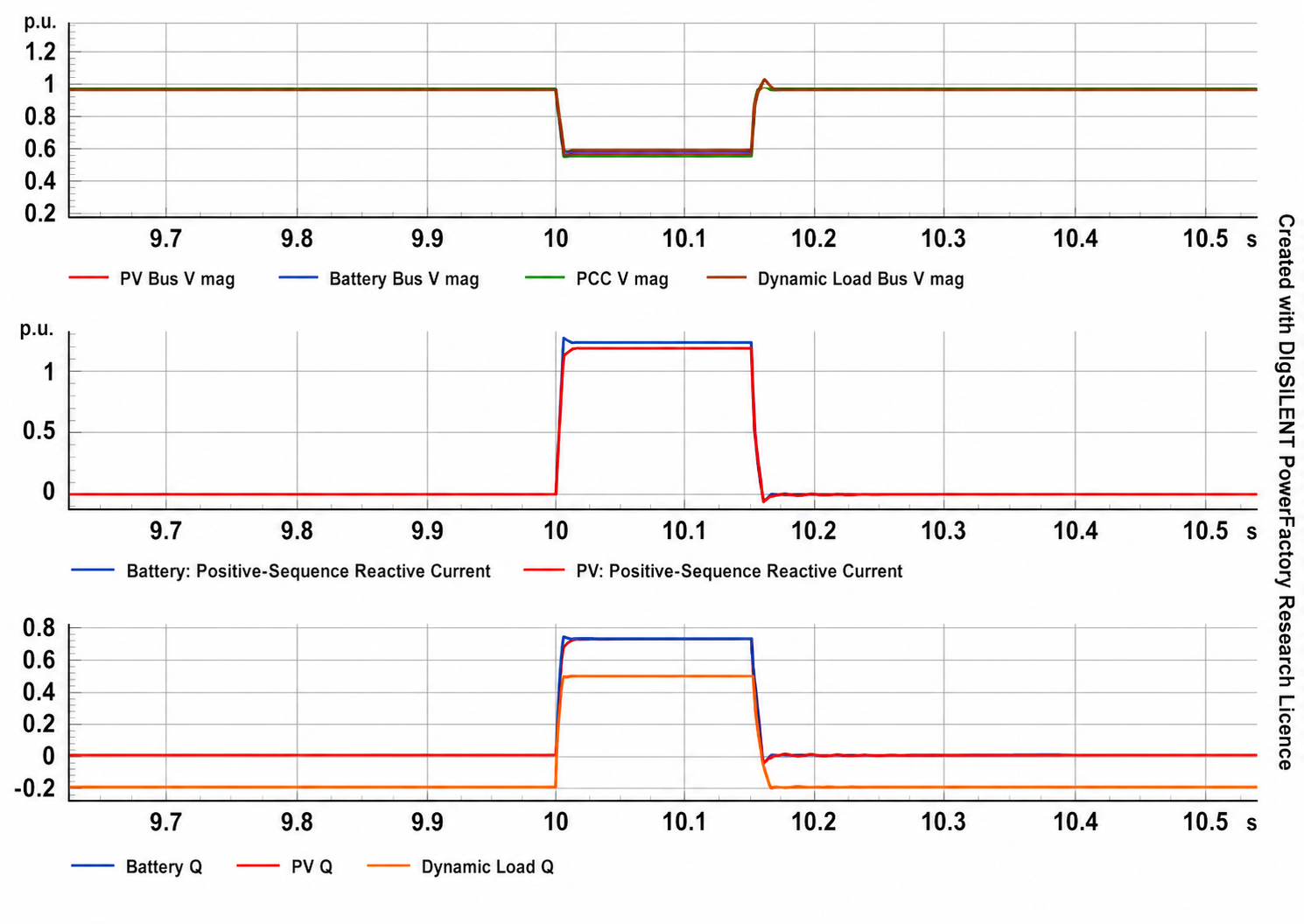}
\caption{Voltage, reactive current, and reactive power in the SCR test, weak grid (SCR $=2.9$).}
\label{fig:scr29_qv}
\end{figure}

A three-phase fault with fault impedance is applied at the PCC with $P=1.0$ pu, for a very strong grid (SCR $=26$, Fig.~\ref{fig:scr26_qv}) and a weak grid (SCR $=2.9$, Fig.~\ref{fig:scr29_qv}). In the strong grid, the voltage dips to about 0.85 pu and recovers within 60 ms, reactive current peaks at about 0.3 pu, and PV and BESS active power falls to 0.75--0.8 pu. In the weak grid, the dip is deeper (0.55--0.78 pu), reactive current reaches about 1.2 pu, and active power drops to 0.4--0.5 pu as the converters approach their current limits; recovery is slower but monotonic. In both cases, the plant remains connected and synchronized, respecting its current limits. Because the dip and the current-limit engagement involve time scales below 10 ms, these results are indicative and must be confirmed with EMT models that include the inner loops.

\subsection{Dynamic Load under NC DCC Tests (D1, D3)} \label{sec:dcc}
In the frequency test (D1), the load remains connected at 49 Hz with constant consumption, consistent with unlimited operation in the 49--51 Hz band. At 47.5 Hz and 51.5 Hz it reduces its demand according to the 5\% droop (to about 0.7 and 0.82 MW, respectively) and disconnects once the permitted operating time, scaled to 2 s in simulation, is exceeded. In the voltage ride-through test (D3), the bus voltage is reduced to 0.2 and 0.4 pu. The load stays connected in both cases; its active power falls from 1.0 MW to about 0.05 and 0.18 MW, respectively, while the positive-sequence reactive current reaches about 0.80 and 0.76 pu. After the dip, the voltage recovers to 1.0 pu without overshoot or oscillation and the load returns to its pre-fault demand. Because D3 involves fast current dynamics, it is, like E3, indicative in RMS.

\subsection{Summary of ENTSO-E Tests}
Table~\ref{tab:ComplianceSummary} summarises the assessment of all eight ENTSO-E tests for the individual units and the co-located system. Tests marked RMS-indicative (E1--E3) are those whose criteria involve sub-10 ms dynamics (Table~\ref{tab:Tests}); for these, the RMS result indicates that the control structure behaves as intended but does not constitute a compliance verdict. During LVRT, operation at the current limit is expected, since converters shall not exceed their rated current.
\begin{table*}[!t]
\centering
\caption{Assessment of ENTSO-E tests for the individual units and the co-located PV--BESS--load system (RMS)}
\label{tab:ComplianceSummary}
\footnotesize
\setlength{\tabcolsep}{4pt}
\begin{tabularx}{\textwidth}{@{}L{4.3cm}L{2.2cm}YY@{}}
\toprule
\textbf{Test} & \textbf{RMS result} & \textbf{Key finding} & \textbf{Limiting factor} \\
\midrule
E1 Phase angle step ($\pm 5^\circ$) & Pass (indicative) & $\Delta i_p \approx 0.29$/0.30 pu (PV/BESS) vs.\ 0.29/0.32 pu calculated; $\geq 50\%$ within 10 ms & EMT needed for 10 ms criteria \\
E2 Small voltage step (0.95 pu) & Pass (indicative) & $i_q$ 0.12--0.15 pu within 10 ms, 90\% within 30 ms, settling 60 ms & EMT needed for 10 ms criteria \\
E3 LVRT (0.2 pu and 0 pu, 150 ms) & Pass (indicative) & $i_q \approx 1.2$ pu (0.2 pu dip) and $\approx 1.5$ pu (solid fault) within 10 ms & Current limit (1.5 pu) \\
E4 LLSG (islanding) & Pass & Autonomous formation, $f \to 49.85$ Hz, $V \approx 0.955$ pu & Requires pre-event headroom \\
E5 RoCoF up ($+4$ Hz/s) & Pass & 10--90\% power within 100 ms & None \\
E6 RoCoF down ($-4$ Hz/s) & Pass & 10--90\% power within 100 ms & Clamped at rated power from 1.0 pu \\
E7 SCR variation (2.9--26) & Pass & Stable, no oscillation; $i_q \approx 1.2$ pu at SCR 2.9 & Deeper dips at SCR 2.9 \\
E8 RoCoF + phase step & Pass & Well damped, synchronised & No critical interaction \\
\bottomrule
\end{tabularx}
\end{table*}

\subsection{Scope and Limitations}
The results constitute an RMS-based pre-compliance assessment, not certification. (i) RMS models omit inner current/voltage loops, network electromagnetic dynamics, and harmonic or sub-synchronous control interactions; tests E1--E3 and D3 therefore require EMT verification, and the toolbox is intended to screen operating points and parameter sets before such studies. (ii) Hardware non-idealities, measurement and communication delays, and protection interactions are not represented, so field or hardware-in-the-loop tests remain necessary. (iii) The test suite covers the ENTSO-E and NC DCC procedures; other events (e.g., unbalanced faults, for which the library VSM has no negative-sequence current control, and fault durations of 0.5--1 s typical of 20 kV networks) are not included. (iv) The model layer is specific to PowerFactory and to the assets studied.

\section{Conclusion}
This paper presented an RMS test bench and a Python--PowerFactory toolbox for checking GFM PV, GFM BESS, and converter-based dynamic loads against the ENTSO-E GFM tests and the ACER NC DCC. The extended models include DC-side dynamics, VSM inertia and droop, virtual impedance, and current limiting. For the co-located system, the RMS-verifiable tests (RoCoF, LLSG, SCR variation) show synthetic inertial response, droop-based power sharing, voltage support, and stable islanded and weak-grid operation for SCR between 2.9 and 26, while the fast tests are flagged for EMT confirmation. Future work will validate the fast tests (E1--E3, D3) with EMT and real-time hardware-in-the-loop simulation, study a BESS-referenced configuration with realistic DC-link energy, and automate parameter tuning.

\section{References}
\bibliographystyle{IEEEtran}
\bibliography{ref}

\end{document}